\RequirePackage{etoolbox}
\csdef{input@path}{%
 {sty/}% cls, sty files
 {img/}% eps files
}%
\csgdef{bibdir}{bib/}% bst, bib files

\documentclass[ba]{imsart}
\pubyear{0000}
\volume{00}
\issue{0}
\doi{0000}
\firstpage{1}
\lastpage{1}

\usepackage{amsthm}
\usepackage{amsmath}
\usepackage{natbib}
\usepackage[colorlinks,citecolor=blue,urlcolor=blue,filecolor=blue,backref=page]{hyperref}
\usepackage{graphicx}
\usepackage{color}
\usepackage[english]{babel}
\usepackage{dsfont}

\startlocaldefs
\numberwithin{equation}{section}
\theoremstyle{plain}

\newcommand{\rmT}{{\rm T}}
\newcommand{\bx}{\mathbf{x}}
\newcommand{\Prob}{{\rm Prob}}
\newcommand{\rme}{{\rm e}}
\newcommand{\rmd}{{\rm d}}
\newcommand{\balpha}{{\boldsymbol\alpha}}
\newcommand{\btheta}{{\boldsymbol\theta}}
\newcommand{\bmu}{{\boldsymbol\mu}}
\newcommand{\mM}{\mathcal{M}}

\newcommand{\mX}{\mathcal{X}}
\endlocaldefs

\begin{document}

\begin{frontmatter}
\title{Covariant priors and model uncertainty}
\runtitle{Covariant priors and model uncertainty}

\begin{aug}
\author{\fnms{Giovanni} \snm{Mana}\thanksref{addr1}\ead[label=e1]{g.mana@inrim.it}}
\and
\author{\fnms{Carlo} \snm{Palmisano}\thanksref{addr2}\ead[label=e2]{palmisan@to.inf.it}}

\runauthor{G.\ Mana and C.\ Palmisano}

\address[addr1]{INRIM -- Istituto Nazionale di Ricerca Metrologica, str.\ delle Cacce 91, 10135 Torino, Italy
    \printead{e1} % print email address of "e1"
}

\address[addr2]{UNITO - Universit\`a di Torino, Dipartimento di Fisica, v.\ P.\ Giuria 1, 10125 Torino, Italy
    \printead{e2}
}
\end{aug}

\begin{abstract}
In the application of Bayesian methods to metrology, pre-data probabilities play a critical role in the estimation of the model uncertainty. Following the observation that distributions form Riemann's manifolds, methods of differential geometry can be applied to ensure covariant priors and uncertainties independent of parameterization. Paradoxes were found in multi-parameter problems and alternatives were developed; but, when different parameters are of interest, covariance may be lost. This paper overviews information geometry, investigates some key paradoxes, and proposes solutions that preserve covariance.
\end{abstract}

% 62C10  	Bayesian problems; characterization of Bayes procedures
% 62F15  	Bayesian inference
% 62F03  	Hypothesis testing
% 62F07  	Ranking and selection
% 68T37  	Reasoning under uncertainty


\end{frontmatter}

\section{Introduction}
In metrology, the Bayes theorem allows the information about a measurand -- jointly delivered by the data and prior knowledge -- to be summarized by probability assignments to its values [\cite{Jaynes:2003,D'Agostini:2003,MacKay:2003,Sivia:2006,Linden:2014}]. This synt
hesis is an essential step to express the measurement uncertainty and to take decisions based on the measurand value.

Including the model in the assessment of the measurement uncertainty requires model selection and averaging [\cite{Dose:2007,Elster:2010,Mana:2012,Toman:2012,Mana:2014,Mana:2015}]; in turn, this requires the model probabilities, which probabilities are proportional to the marginal likelihood, also termed model evidence. Reporting the posterior distribution -- or a summary, like point and interval estimates -- and marginal likelihood -- to carry out model selection or to estimate the model uncertainty -- is part of the data analysis. To make a meaningful posterior distribution and uncertainty assessment, the prior density must be covariant; that is, the prior distributions of different parameterizations must be obtained by transformations of variables. Furthermore, it is necessary that the prior densities are proper.

If a preferred parameterization exists, the prior density of any other parameterization is obtained by transformations of variables. We will not examine how to include explicitly-stated information into prior probability distributions. The idea is to find the maximum-entropy prior subject to the given constraints; discussions can be found in \cite{Jaynes:2003,D'Agostini:2003,MacKay:2003,Sivia:2006,Linden:2014}. If the only information is the statistical model explaining the data, since parametric models form Riemann's manifolds, priors derived from the manifold metric -- for instance, but not necessarily, the Jeffreys ones -- ensure the sought covariance, while finite manifold volumes ensure proper priors. The prior density is a part of the model: non-covariant priors go with different metrics and, thus, with different models. To avoid that metrics affect model selection and uncertainty, the prior determining-rule must ensure the metric invariance, that is, different parametric models explaining the same data must have the same metric.

Differential geometry was introduced in statistic by Rao [\cite{Rao:1945}], who triggered the development of what is now known as information geometry [\cite{Amari:2007,Arwini:2008}]. The same ideas led Jeffreys to determine covariant priors from the information metric [\cite{Jeffreys:1946,Jeffreys:1998}]. The sections from 2 to 5 give to metrologists an overview of the methods of information geometry as applied to encode the model information into covariant distributions of the parameters and to overcome the feeling of lack of objectivity when carrying out Bayesian inferences. More extensive reviews are in \cite{Kass:1989,Kass:1996,Costa:2014}.

When the Jeffreys rule is used in multi-parameter problems, paradoxes and inconsistencies were observed and alternatives were developed [\cite{Bernardo:1979,Kass:1996,Datta:1995}]. A milestone are the Berger and Bernardo reference priors, that divide the model parameters in parameters of interest -- i.e., the measurands -- and nuisance parameters [\cite{Berger:1992,Berger:1992a,Bernardo:2005,Berger:2008,Berger:2009,Berger:2015,Bodnar:2015}]. Roughly, they maximise the expected information gain -- the Kullback-Leibler divergence between the prior and posterior distributions -- in the limit of infinite measurement repetitions. For regular models where asymptotic normality of the posterior distribution holds, if there aren't nuisance parameters, the reference priors are the Jeffreys ones [\cite{Ghosh:2011}]. However, if there are nuisance parameters, the reference priors depend on the measurands and covariance is lost [\cite{Datta:1996}].

The section 6 investigates some of the key inconsistencies and paradoxes reported in the literature. The aim is not to challenge the existence of the inconsistencies or the Berger and Bernardo reference priors, but to show that solutions that save the prior covariance exist. Firstly, we examines why, when a uniform prior is used to infer the measurands' values from repeated Gaussian measurements, there is no greater uncertainty with many being estimated than with only one [\cite{Jeffreys:1946,Jeffreys:1998}]. Next, since when using covariant priors the expected frequencies depend on the cell number, it considers the problem of determining how many outcomes to include in a multinomial model [\cite{Bernardo:1989,Berger:2015}]. Eventually, after recognising the diversity and uncertainty of the underlying models, it proposes covariant solutions to the Stein [\cite{Stein:1959,Attivissimo:2012,Samworth:2012,Carobbi:2014}] and Neyman-Scott paradoxes [\cite{Neyman-Scott:1948}].

\section{Background}
This section introduces Bayesian data analysis, information geometry, and model selection. It does not aim at giving an exhaustive review, but it is a courtesy to metrologists and readers who do not master these fields.

Let us consider the measurement of a scalar quantity $\alpha$ and the sampling distribution $p(x|\alpha)$ of the measurement result $x$ if the measurand value is $\alpha$, where we use the same symbol both to indicate random variables and to label the space of the their values. The extension to many measurands, the presence of nuisance parameters, and multivariate distributions does not require new concepts and will not be examined in detail. The distribution family $\mathcal{M}=\{p(x|\alpha)\,:\, \alpha\in\mathds{R}\}$, whose elements are parameterized by $\alpha$, is the parametric model of the data. It is worth noting that $\alpha$ is a coordinate labelling the $\mM$'s elements.

The post-data distribution of the measurand values, which updates the information available prior the measurement and synthesized by the prior density $\pi(\alpha|\mM)$, is
\begin{equation}\label{bayes}
 p(\alpha|x,\mM) = \frac{L(\alpha|x)\pi(\alpha|\mM)}{Z(x|\mM)} ,
\end{equation}
where $L(\alpha|x)=p(x|\alpha)$ is the likelihood of the model parameters -- which is the sampling distribution itself, now, read as a function of the model parameters given the data. The normalizing constant, which is termed marginal likelihood or evidence,
\begin{equation}\label{evidence}
 Z(x|\mM) = \int_{-\infty}^{+\infty}\!\! L(\alpha|x)\pi(\alpha|\mM)\, \rmd\alpha
\end{equation}
is the probability distribution of the data given the model. As such, it must be independent of the model parameters. Section \ref{model-selection} will show that (\ref{evidence}) is crucial in the evaluation of how much the data support the explanation $\mM$. The symbols $\pi(\_|\mM)$  and $p(\_|\_)$ will be used to indicate prior and posterior probability densities and the relevant conditioning, not given functions of the arguments.

If no information is available, $\pi(\alpha|\mM)$ must not deliver information about $\alpha$. When it is a continuous variable, we can consider the continuous limit of $\pi_n=\Prob(\alpha\in\Delta_n)$, where $\Delta_n$ is the length of the $n$-th interval in which the measurand domain has been subdivided and $1\le n\le N$. Hence, $\pi(\alpha|\mM)=\lim_{N\rightarrow\infty}\pi_n/\Delta_n$, where $\max(\Delta_n)\rightarrow 0$. Since, in the class of the finitely discrete ones, the uninformative distribution is $\pi_n=1/N$, the sought prior density is seemingly found.

However, different limit procedures originate different distributions and nothing indicates what should be preferred. For instance, if $\Delta_n=A/N$, where $A$ is the range of the measurand values, $\pi(\alpha|\mM)=1/A$ will follow. If $\Delta_n=1/[\mu(\alpha)N]$, where $\mu(\alpha)$ is any probability measure, the prior density is $\pi(\alpha|\mM)=\mu(\alpha)$. In the same way, if the measurand is changed to $\beta(\alpha)$, the $\beta$'s distribution,
\begin{equation}
 \pi'(\beta|\mM) = \pi\big[\alpha(\beta)|\mM\big] \left| \frac{\rmd \alpha}{\rmd \beta} \right| ,
\end{equation}
where $\pi(\alpha|\mM)$ represents the absence of information and $\alpha(\beta)$ is the inverse transformation, is, in general, a different function. For instance, if $\beta=\alpha^2$, the transformed distribution is $\pi'(\beta|\mM) = \pi(\sqrt{\beta}|\mM)/\sqrt{4\beta}$. Also in this case, since we are as ignorant about $\beta$ as about $\alpha$, nothing indicates what prior distribution is to be used.

These difficulties arise because we assumed that no information is available. But this is not true: as the conditioning on $\mM$ indicates, $\alpha$ labels the elements of the model explaining the data. It is a way to get a representation, to write formulae explicitly, but, in principle, its choice is arbitrary. Hence, a measurand change corresponds to a mere coordinate change. A parameter-free way to express ignorance is to require that the elements of $\mM$ are equiprobable. Hence, we must supply a metric; this can be done in a natural way, as it will be presently shown.

\section{Information geometry}
This section overviews the methods of information geometry as applied to encode by covariant rules the information delivered by the data models into prior densities. To reduce the algebra to a minimum, we consider only the single measurand case; full treatments can be found in  [\cite{Amari:2007,Arwini:2008}].

Let us introduce the probability amplitudes $\{\psi(x|\alpha)=\sqrt{p(x|\alpha)} \,:\, p(x|\alpha)\in\mathcal{M} \}$. Since they are a subset of the $\mathcal{L}^2$ space of the square-integrable functions, $\mathcal{M}$ inherits the 2-norm metric
\begin{equation}\label{distance}
 D(\alpha_2, \alpha_1) = \int_{-\infty}^{+\infty}\!\! |\psi(x|\alpha_2)-\psi(x|\alpha_1)|^2\, \rmd x
\end{equation}
induced by the $\mathcal{L}^2$ scalar product. To obtain the metric tensor, we observe that the line element is
\begin{equation}\label{line}
 \rmd s^2 = D(\alpha+\rmd\alpha, \alpha) =
 \left( \int_{-\infty}^{+\infty}\!\! |\partial_\alpha\psi(x|\alpha)|^2\, \rmd x \right) \rmd\alpha^2 .
\end{equation}
Next, since
\begin{equation}\label{line:1}
 \partial_\alpha\psi = \frac{\psi\partial_\alpha\ln(\psi^2)}{2} ,
\end{equation}
where the units have been chosen in such a way to make the $x$ variable dimensionless, we can rewrite (\ref{line}) as
\begin{eqnarray}
 4\rmd s^2 &= &\left( \int_{-\infty}^{+\infty}\!\! \big| \partial_\alpha\ln[p(x|\alpha)] \big|^2 p(x|\alpha)\, \rmd x \right)
 \rmd\alpha^2 \nonumber \\
      &= &\left\langle \big| \partial_\alpha\ln(L) \big|^2 \right\rangle \rmd\alpha^2
       =  J(L;\alpha)\, \rmd\alpha^2 ,
\end{eqnarray}
where $L=L(\alpha|x)$ is the likelihood, the angle brackets indicate the average and $J(L;\alpha)$ is the Fisher information, which is proportional to the metric tensor. It is worth noting that, if $ \partial^2_\alpha\ln(L)$ exists, the Fisher information can also be written as
\begin{equation}
 J(L;\alpha) = -\left\langle \partial^2_\alpha\ln(L) \right\rangle\ .
\end{equation}

The metric tensor is related to the Kullback-Leibler divergence,
\begin{equation}\label{KL}
 D_{KL}(\alpha_2, \alpha) = \int_{-\infty}^{+\infty}\!\! \ln \left[\frac{p(x|\alpha_2)}{p(x|\alpha)}\right] p(x|\alpha_2) \, \rmd x ,
\end{equation}
as follows. By expanding (\ref{KL}) in series of $\rmd\alpha=\alpha_2-\alpha$ and taking the normalization of $p(x|\alpha)$ into account, we obtain
\begin{equation}
 D_{KL}(\alpha+\rmd\alpha, \alpha) = \frac{1}{2}J(L;\alpha)\, \rmd\alpha^2 ,
\end{equation}
up to higher order terms. Therefore, $\rmd s^2$ measures the information gain when $p(x|\alpha+d\alpha)$ updates $p(x|\alpha)$.

When there are $p$ parameters, so that $\balpha=[\alpha_1, ...\, \alpha_p]^{\rm T}$ is a $p\times 1$ matrix, the Fisher information is the $p\times p$  matrix $\mathds{J}(L;\balpha) = -\langle \nabla^{\rm T}_\alpha \nabla_\alpha \ln(L) \rangle = \langle \nabla^{\rm T}_\alpha \ln(L) \nabla_\alpha \ln(L) \rangle$, where $\nabla_\alpha = [\partial_1, \partial_2, ...\, \partial_p]$, and
\begin{equation}
 \rmd s^2 \propto \rmd\balpha^{\rm T} \mathds{J}(L;\balpha)\, \rmd\balpha .
\end{equation}

\section{Covariant priors}
The prior for the model parameters must embed the parameter meaning in the problem at hand, which is defined by the model that is assumed to explain the data [\cite{Linden:2014}]. Therefore, it makes sense to chose a prior density that is the uniform measure of $\mM$ with respect to its metric, that is,
\begin{equation}
 \Prob\left[ p(x|\balpha) \in \rmd V_{\balpha} \right] \propto \rmd V_{\balpha} ,
\end{equation}
where, by using the $\alpha$ parameterization, the volume element is
\begin{equation}\label{Vel}
 \rmd V_{\balpha} = \sqrt{\det(\mathds{J})}\, \rmd\alpha_1 \,...\, \rmd\alpha_p .
\end{equation}
This induces the Jeffreys probability density [\cite{Jeffreys:1946,Jeffreys:1998}]
\begin{equation}\label{Jeffreys}
 \pi_J(\balpha|\mM) \propto \sqrt{\det(\mathds{J})} ,
\end{equation}
which is the continuous limit of a discrete distribution defined over a lattice where the node spacings ensure that the same information is gained when the sampling distribution labelled by a node updates those identified by its neighbourhoods.

The relevance of the Jeffreys rule in metrology and in expressing uncertainties in measurements resides in the metric invariance -- that is, the same metric is assumed for all the models that explain the data -- and prior covariance under one-to-one coordinate transformations -- that is, under reparameterization of the sampling distribution. Covariance makes the post-data distribution (\ref{bayes}) consistent with transformations of the model parameters. In fact, the left-hand side of (\ref{bayes}) transforms according to the usual change-of-variable rule. What happens to the right-hand side is that the transformation Jacobian combines with $\mathds{J}(L,\balpha)$ to give the Fisher information about the new variables. This is a consequence of the invariance of the volume element (\ref{Vel}).

Here is the explicit proof in the single-parameter case. Let $\beta(\alpha)$ be a one-to-one coordinate transformation, i.e., a reparameterization of the sampling distribution. By application of (\ref{bayes}), the post-data distribution of the $\beta$ measurand is
\begin{equation}\label{beta}
 p'(\beta|x,\mM) \propto \frac{L'(\beta|x) \sqrt{J(L';\beta)}}{Z'(x|\mM)} ,
\end{equation}
where the normalization factor of $\sqrt{J(L';\beta)}$ has been omitted, $L'(\beta|x)=L\big[\alpha(\beta)|x\big]$ is the reparameterized likelihood, $\alpha(\beta)$ is the inverse transformation, and $\pi'_J(\beta) \propto \sqrt{J(L';\beta)}$ is the Jeffreys prior of $\beta$. Since
\begin{eqnarray}\nonumber\label{normJ}
 \sqrt{J(L;\alpha)} \left| \frac{\rmd \alpha}{\rmd \beta} \right|
 &= &\sqrt{ \left\langle \left| \frac{\rmd \alpha}{\rmd \beta} \partial_\alpha\ln(L) \right|^2 \right\rangle } \nonumber \\
 &= &\sqrt{ \left\langle \left| \partial_\beta\ln(L') \right|^2 \right\rangle } = \sqrt{J(L';\beta)} ,
\end{eqnarray}
by applying the change of variable rule to the $p(\alpha|x,\mM)$ post-data distribution of $\alpha$ in (\ref{bayes}), we obtain
\begin{eqnarray}
 p'(\beta|x,\mM)
 &= &p\big[ \alpha(\beta)|x,\mM \big] \left| \frac{\rmd \alpha}{\rmd \beta} \right| \nonumber \\
 &\propto &\frac{L\big[\alpha(\beta)|x\big] \sqrt{J\big[L;\alpha(\beta)\big]} \left| \displaystyle\frac{\rmd \alpha}{\rmd \beta} \right|}{Z(x|\mM)} \nonumber \\
 &= &\frac{L'(\beta|x) \sqrt{J(L';\beta)}}{Z(x|\mM)} . \label{beta2}
\end{eqnarray}
Eventually, it is easy to prove that (\ref{bayes}) and (\ref{beta2}) deliver the same marginal likelihood. In fact,
\begin{eqnarray}
 Z(x|\mM) &\propto &\int_{-\infty}^{+\infty}\!\! L(\alpha|x)\sqrt{J(L;\alpha)}\, \rmd\alpha \nonumber \\
       &= &\int_{-\infty}^{+\infty}\!\! L'(\beta|x)\sqrt{J(L';\beta)}\, \rmd\beta \propto Z'(x|\mM) ,
\end{eqnarray}
where the same normalization factor -- by virtue of (\ref{normJ}) -- of $\sqrt{J(L;\alpha)}$ and $\sqrt{J(L';\beta)}$ has been left out. This identity expresses that the marginal distribution of the data is independent of the representation of the model elements.

\section{Model selection and uncertainty}\label{model-selection}
Let the data be explained by a set of mutually exclusive models that are hyper-parameterized by $k$, which can be both discrete (e.g., the cell label of a multinomial model) or continuous (e.g., the domain boundary of the mean of a Gaussian models). Assuming that the $\{\mM_k\}$ set is complete, by application of the Bayes theorem, the posterior odds on $\mM_k$ explaining the data are [\cite{MacKay:2003,Sivia:2006,Linden:2014}]
\begin{equation}\label{PM1}
 \Prob(k|x) = \frac{Z(x|k)\pi(k)}{\sum_k Z(x|k) \pi(k)} ,
\end{equation}
where $\pi(k)$ is the prior probability of $\mM_k$ and, when, $k$ is continuous, an integration substitutes for the sum. In (\ref{PM1}), the marginal likelihood of a previous Bayesian analysis, $Z(x|k)$, is the model likelihood. It is worth noting that, since it is the sampling distributions of the data given $\mM_k$, $Z(x|k)$ must be independent of the model parameters $\alpha$; in turn, the prior density $\pi(\alpha|k)$ must be covariant.

The marginal likelihood is proportional to the power of the model to explain the data -- the higher is the fitness, the greater $Z(x|k)$ -- but inversely proportional to the volume of the parameter space -- the greater is the model freedom, the lesser $Z(x|k)$. This characteristic of $Z(x|k)$, known as the Ockham's razor [\cite{MacKay:2003,Sivia:2006,Linden:2014}], penalizes the models that, by adjusting the model parameters, have a greater freedom to explain the data. Hence, improper priors, which always correspond to infinite parameter volumes and freedom, make the probability of observing the data set null and, consequently, are never supported; also if they correspond to proper posterior distributions. This issue will be further examined in the next section.

Ensuring that marginal likelihoods do not depend on the $\mM_k$ parameterization makes it possible to calculate the model uncertainty and, in turn, it makes it possible to include the model uncertainty in the error budget [\cite{Clyde:2004}]. In the same way as $p(\alpha|x,k)$ expresses the uncertainty of the measurand value -- provided that $\mM_k$ explains the data -- $\Prob(k|x)$ expresses the model uncertainty. By combining the measurand and model distributions and by marginalising over the models, the total uncertainty is expressed by
\begin{equation}\label{mave}
 p(\alpha|x) = \sum_k p(\alpha|x,k)\Prob(k|x) ,
\end{equation}
where, if $k$ is continuous, an integration substitutes for the sum.

\section{Paradox analyses}
In the following, to avoid increasing too much the length of the paper, intermediate calculations are omitted. All were carried out with the help of Mathematica$^\text{\textregistered}$ [\cite{Mathematica:9}].

\subsection{Gaussian model with standardized mean}
To exemplify the issues of the use of non-covariant priors, let us consider $n$ independent samples $\bx = [x_1,x_2,\, ... \,x_n]^{\rm T}$ from the Gaussian model $\mM_\mu= \{N(x_i|\mu,\sigma) \,:\, (\mu,\sigma)\in\mathds{R}\times\mathds{R}^+\}$, where both the mean $\mu$ and standard deviation $\sigma$ are unknown. To simplify the analysis, we set the data offset and measurement unit in such a way that the sample mean and (biased) standard deviation are zero and one, respectively. Hence, the sampling distribution of $\bx$ is
\begin{equation}\label{Gauss-model}
 p(\bx|\mu,\sigma) = \frac{1}{(\sqrt{2\pi}\,\sigma)^n} \exp{\left[-\frac{n(1+\mu^2)}{2\sigma^2}\right]} .
\end{equation}
The Berger and Bernardo reference prior [\cite{Bernardo:1979,Berger:2015}] to make inferences about the $\{\mu,\sigma\}$ pair is
\begin{equation}\label{pr1}
 \pi_{\rm BB}(\mu,\sigma|\mM_\mu) \propto 1/\sigma ,
\end{equation}
which results in the marginal likelihood
\begin{equation}\label{Z1}
 Z(\bx|\mM_\mu) = \int_0^\infty\!\! \int_{-\infty}^{+\infty}\!\! p(\bx|\mu,\sigma) \pi_{\rm BB}(\mu,\sigma|\mM_\mu)\, \rmd \mu\, \rmd\sigma
 \propto \frac{\Gamma(n/2-1/2)}{2 \sqrt{n^n \pi^{n-1}}} ,
\end{equation}
where $\Gamma(z)$ is the Euler gamma function, and posterior distribution
\begin{equation}\label{post1}
 p(\mu,\sigma|\bx,\mM_\mu) = \frac{ \sqrt{n^n} }{\sqrt{2^{n-2}\pi}\,\sigma^{n+1}\Gamma(n/2-1/2)}
 \exp{\left[-\displaystyle\frac{n(1+\mu^2)}{2\sigma^2}\right]} .
\end{equation}
The proportionality sign in (\ref{Z1}) stresses that, since $\pi_{\rm BB}(\mu,\sigma|\mM_\mu)$ is improper; hence, the marginal likelihood is defined only up to an undefined factor. Actually, the normalizing factor of $\pi_{\rm BB}(\mu,\sigma|\mM_\mu)$ is infinite; therefore, $Z(\bx|\mM_\mu)$ is null. This means that the prior (\ref{pr1}), more precisely its support, is falsified by the data. This problem will be carefully analysed in the following.

If the measurands are the standardized mean $\lambda=\mu/\sigma$ and $\sigma$, the data model is re-parameterized as $\mM_\lambda= \{ N(x_i|\sigma\lambda,\sigma) \,:\, (\lambda,\sigma)\in\mathds{R}\times\mathds{R}^+\}$. Hence, the sampling distribution (\ref{Gauss-model}) is reparameterized as
\begin{equation}
 p(\bx|\lambda,\sigma) = \frac{1}{(\sqrt{2\pi}\,\sigma)^n} \exp{\left[-\frac{n(1+\lambda^2\sigma^2)}{2\sigma^2}\right]} .
\end{equation}
The Berger and Bernardo reference prior [\cite{Bernardo:1979,Berger:2015}] to make inferences about the $\{\lambda,\sigma\}$ pair is
\begin{equation}\label{pr2}
 \pi_{\rm BB}(\lambda,\sigma|\mM_\lambda) \propto \frac{1}{\sqrt{2 + \lambda^2}\,\sigma} ,
\end{equation}
which results in the marginal likelihood [\cite{Mathematica:9}]
\begin{eqnarray}\label{Z2}\nonumber
 Z(\bx|\mM_\lambda) &= &\int_0^\infty\! \int_{-\infty}^{+\infty}\!\!
 p(\bx|\lambda,\sigma) \pi_{\rm BB}(\lambda,\sigma|\mM_\mu)\, \rmd \lambda\, \rmd\sigma \\
 &\propto &\frac{K_0(n/2)\Gamma(n/2+1)\rme^{n/2}}{n\sqrt{(n\pi)^n}} ,
\end{eqnarray}
where $K_0(z)$ is the modified Bessel function of the second kind, and posterior distribution
\begin{eqnarray}\label{post2}\nonumber
 p(\lambda,\sigma|\bx,\mM_\lambda) &= &\frac{ n\sqrt{n^n} }
 {\sqrt{2^n(2+\lambda^2)}\,\sigma^{n+1}K_0(n/2)\Gamma(n/2+1)} \\
 & &\times \exp{\left[-\displaystyle\frac{n(1+(1+\lambda)\sigma^2)} {2\sigma^2}\right]} .
\end{eqnarray}
Also in this case, since (\ref{pr2}) is improper, the proportionality sign in (\ref{Z2}) stresses that $Z(\bx|\mM_\lambda)$ is defined only up to an undefined factor; actually, in the limit when the prior support is $\mathds{R}\times\mathds{R}^+$ it is zero.

\begin{figure}\centering
\includegraphics[width=7.5cm]{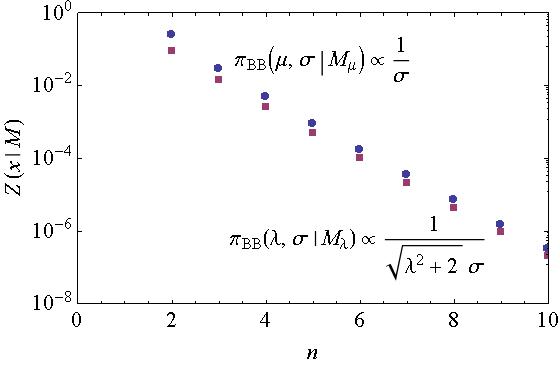}
\caption{Marginal likelihoods of $n$ independent data drawn from the same Gaussian distribution and the Berger and Bernardo reference priors (\ref{pr1}), blue, and (\ref{pr2}), magenta.} \label{ref-prior}
\end{figure}

The priors (\ref{pr1}) and (\ref{pr2}) are tailored to the different measurands, but they are not covariant. This originates a twofold difficulty. Firstly, notwithstanding they are conditioned to the same sampling distribution, if we insist on such an interpretation, the data probabilities derived from (\ref{Z1}) and (\ref{Z2}) are different, as also shown in Fig. \ref{ref-prior}. Secondly, after having the distribution (\ref{post1}), one can legitimately use it to calculate $p(\lambda,\sigma|\bx,\mM_\lambda)$ by changing the variables from $\{\mu,\sigma\}$ to $\{\lambda,\sigma\}$. Also in this case, notwithstanding it is conditioned to the same data, sampling distribution, and prior information, the result is different from (\ref{post2}).

The paradox is explained by observing that the prior densities (\ref{pr1}) and (\ref{pr2}) embed different information. Since they have different metrics, $\mM_\lambda$ is not a re-parameterization of $\mM_\mu$ and, therefore, the two analyses rely on different models. Bayesian model selection might be used to choose between the competing models. However, one must be aware that, in this case, the selection will act on the metrics of the sampling distributions. A second issue is that, since (\ref{pr1}) and (\ref{pr2}) are improper, the marginal likelihood (\ref{Z1}) and (\ref{Z2}) are defined only apart undefined proportionality factors. This makes model selection meaningless.

\subsection{Multinormal distribution}
Jeffreys [\cite{Jeffreys:1946,Jeffreys:1998}] modified the rule (\ref{Jeffreys}) because, as we can understand, when applied to independent Gaussian measurements of many measurands having common variance, the degrees of freedom of the posterior distributions of any measurand subset, depend only on the number of measurements, regardless of the measurand number. This implies that the variance of any measurand is the same, no matter how many are estimated. In this section, we will examine this issue in detail.

\subsubsection{Problem statement.}
Let $\mathds{X} = \{x_{ij}\}$ be $n$ realizations of $m$ independent normal variables with unknown means $\mu_i$ and variance $\sigma^2$, for $i=1,2,\, ...\,m$ and $j=1,2,\, ...\, n$. The sampling distribution of $\mathds{X}$ is
\begin{equation}
 p(\mathds{X}|\bmu,\sigma)=\prod_{j=1}^n N_m(\bx_j|\bmu,\sigma^2 \mathds{I}_m) ,
\end{equation}
where  $\bmu =[\mu_1,\mu_2, ...\, \mu_m]^\rmT$,  $\bx_j =[x_1,x_2, ...\, x_m]_j^\rmT$, and $\mathds{I}_m$ is the $m\times m$ identity matrix. The Jeffreys prior of $\{\bmu,\sigma\}$ is
\begin{equation}\label{prior:Np}
  \pi_J(\bmu,\sigma|\mM) = \frac{m\sigma_0^m}{V_\mu \sigma^{m+1}}
\end{equation}
and $\mM$ is the data model where $V_\mu$ is the volume of the $\bmu$ subspace and $\sigma > \sigma_0$. In the following, the (\ref{prior:Np}) support will be identified with $\mathds{R}^m \times \mathds{R}^+$. However, it is unknown and should be chosen by model selection. This problem will be examined in section \ref{Stein}.

The joint post-data distribution of $\bmu$ and $\sigma$ is
\begin{eqnarray}\label{post-multi}\nonumber
 p(\bmu,\sigma|\mathds{X},\mM)
 &= &\frac{p(\mathds{X}|\bmu,\sigma)}{Z(\mathds{X}|\mM)}\frac{m\sigma_0^m}{V_\mu \sigma^{m+1}} \\
 &= &\frac{1}{Z(\mathds{X}|\mM)}
 \frac{\exp\left(\displaystyle-\frac{mn\overline{s^2}}{2\sigma^2}\right) \exp\left(\displaystyle-\frac{n|\overline{\bx}-\bmu|^2}{2\sigma^2}\right)}
                        { \sqrt{(2\pi)^{mn}}\, \sigma^{mn} } \frac{m\sigma_0^m}{V_\mu \sigma^{m+1}} ,
\end{eqnarray}
where, if the integration domain is extended to $\mathds{R}^m \times \mathds{R}^+$,
\begin{eqnarray}\label{Z-multi}\nonumber
 Z(\mathds{X}|\mM)
 &= &\int_{\sigma_0}^\infty \int_{V_\mu} p(\mathds{X}|\bmu,\sigma) \pi_J(\bmu,\sigma|\mM)\, \rmd\bmu\, \rmd\sigma \\
 &\approx &\frac{\sqrt{(2m)^m}\, \Gamma(mn/2)}{2\sqrt{(mn)^{m(n+1)}\pi^{m(n-1)}}\,\overline{s}^{mn}} \frac{m\sigma_0^m}{V_\mu}
\end{eqnarray}
is the marginal likelihood, $\Gamma(z)$ is the Euler gamma function, $\overline{\bx}  =[\overline{x}_1,\overline{x}_2, ...\, \overline{x}_m]^\rmT$, $\overline{x}_i$ and $s_i^2$ are the mean and (biased) variance of the data set $\{x_{i1}, x_{i2}, ...\, x_{in}\}$, and $\overline{s^2} = (1/m)\sum_{i=1}^m s_i^2$ is the pooled variance. Although the $m\sigma_0^m/V_\mu$ factor of (\ref{Z-multi}) cancels and (\ref{post-multi}) is proper, when $\sigma_0\rightarrow 0$ or $V_\mu \rightarrow \infty$, the data probability $Z(\mathds{X}|\mM)$ is null.

The joint post-data distribution of any subset $\bmu'=[\mu_1, \mu_2, ...\, \mu_q]^\rmT$ of $q\le m$ means is the $q$-variate Student distribution having location $\overline{\bx'}  =[\overline{x}_1,\overline{x}_2, ...\, \overline{x}_q]^\rmT$, scale matrix $\overline{s^2}\mathds{I}_q/n$, and $mn$ degrees of freedom
\begin{eqnarray}\label{student}\nonumber
 t_{mn}( \bmu'|\mathds{X},\mM )
 &= &\int_{V_\mu}\!\! \int_{\sigma_0}^\infty\!\! p(\bmu,\sigma|\mathds{X},\mM)\, \rmd\sigma\, \rmd\mu_{q+1}\,...\,\rmd\mu_m \\
 &\approx &\frac{\Gamma[(mn+q)/2](1+t^2)^{-(mn+q)/2} }{\sqrt{(m\pi)^q} \Gamma(mn/2) \overline{s}^q} ,
\end{eqnarray}
where integration domain is again extended to $\mathds{R}^m \times \mathds{R}^+$ and $t^2=|\overline{\bx'}-\bmu'|^2/(m\overline{s^2})$. The mean and variance-covariance matrix of $\bmu'$ are
\begin{subequations}\begin{equation}\label{t_mean}
 {\rm E}( \bmu'|\mathds{X},\mM ) = \overline{\bx'}
\end{equation}
and
\begin{equation}\label{t_var}
 \Sigma^2( \bmu'|\mathds{X},\mM ) = \frac{m \overline{s^2}}{mn-2} \mathds{I}_q ,
\end{equation}\end{subequations}
where $nm>2$ and $\mathds{I}_q$ is the $q\times q$ identity matrix.

If $m = q = 1$, $t=(\mu-\overline{x})/\overline{s}$ is a Student variable having $n$ degrees of freedom. However, from a frequentist viewpoint, it is a Student variable having $n-1$ degrees of freedom. Furthermore, the variance-covariance matrix (\ref{t_var}) is independent of the number $q$ of estimated means. This implies that, for given sample means and pooled variance from $mn$ observations, there would be no greater uncertainty with $q$ means being estimated than with only one. In particular, the variance of any measurand is $\sigma_\mu^2 = m\overline{s^2}/(mn-2)$, no matter how many are estimated.

\subsubsection{Proposed solution}
The degrees of freedom discrepancy is explained by observing that the Bayesian posterior (\ref{student}) assigns probabilities to the winning elements of a measurands population associated with the same $\mathds{X}$ sample. Contrary, the frequentist distribution assigns probabilities to the $t$-statistic of different $\mathds{X}$ samples given the same measurands. Therefore, there is no reason to expect that the two distributions are the same.

\begin{figure}\centering
\includegraphics[width=7.5cm]{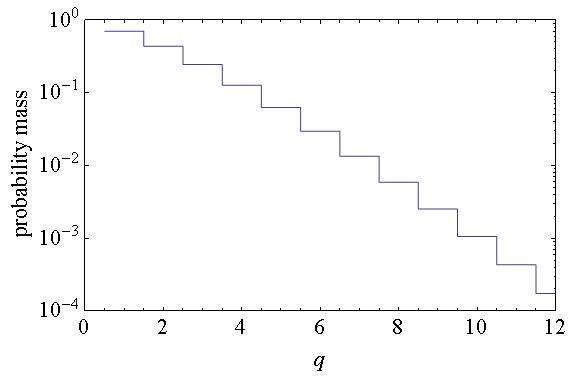}
\caption{Probability that a $q$-subsets of $m$ measurands is in the one standard-deviation credible region. The data are $n$ independent and identically Gaussian measures of each measurand. It is worth noting that the probability depends only on $q$ and the $mn$ degrees of freedom, which were set to 12.} \label{fig01}
\end{figure}

The independence of (\ref{t_var}) from $q$ does not mean that the uncertainty is independent of the measurand number. In fact, the one standard-deviation credible region of $\bmu'$ is a $q$-ball having $\sigma_\mu$ radius. The probability that $\bmu'$  is in this region is [\cite{Mathematica:9}]
\begin{eqnarray}
 \lefteqn{ \int_\Omega \int_0^{\sigma_\mu} t^{q-1} t_{mn} ( \bmu'|\mathds{X} )\, \rmd t \, \rmd\Omega = } \nonumber\\
 && \frac{\Gamma\big[(mn+q)/2\big]} {\sqrt{(mn-2)^q}\, \Gamma(mn/2)}\, _2F_1 \big[ q/2, (mn+q)/2, (q+2)/2, 1/(2-mn) \big] ,
\end{eqnarray}
where $\rmd\Omega$ is for the $m-1$ angular factors and $_2F_1 (a,b,c;z)$ is the hypergeometric function. As shown in Fig.\ \ref{fig01}, as the number of measurands increases, the probability decreases from the maximum -- the probability that $\mu_i$ is in the $[-\sigma_\mu,+\sigma_\mu]$ interval -- to zero. Therefore, despite the variance is the same, there is a greater uncertainty with $q$ measurands being simultaneously estimated than with only one.

\subsection{Multinomial paradox}
We now consider the problem of determining how many outcomes to include in a multinomial model. When using the Jeffreys rule, the expected outcome frequencies depend on the number of cells. In particular, frequencies depend on the number of void cells and tend to zero as they tend to the infinity. Since one has the option of adding outcomes, a paradox arises when the observations are explained by models including an arbitrary number of cells, where no event is observed [\cite{Bernardo:1989,Berger:2015}].

\subsubsection{Problem statement}
Suppose $\bx = \{x_1, x_2,\, ...\, x_m\}$, where $x_i$ are positive integers, is a realization of multinomial variable. Hence, $\bx \sim {\rm{Mu}}(\bx|m,\btheta)$, where $m$ is the cell number, $\btheta=\{\theta_1, \theta_2,\, ... \theta_m\}$ are frequency-of-occurrence, $0\le \theta_i \le 1$, $\sum_{i=1}^m \theta_i = 1$,
\begin{equation}
  {\rm{Mu}}(\bx|m,\btheta) = \frac{n! \prod_{i=1}^m \theta_i^{x_i}}{\prod_{i=1}^m x_i!} ,
\end{equation}
and $n=\sum_{i=1}^m x_i$ is the sample size. The Jeffreys' prior of $\btheta$ is the Dirichlet distribution
\begin{equation}\label{Mu_prior}
 \pi_J(\btheta|m) = \frac{\Gamma(m/2)} {\sqrt{\pi^m} \prod_{i=1}^m \sqrt{\theta_i}} ,
\end{equation}
where $\Gamma(z)$ is the Euler gamma function. The posterior distribution of $\btheta$ given the counts $x_i$ and the $m$-cell explaining model is
\begin{equation}
 p(\btheta|\bx,m) = \frac{{\rm{Mu}}(\bx|m,\btheta) \pi_J(\btheta|m)}{Z(\bx|m)} =
 \frac{\Gamma(n+m/2)\prod_{i=1}^m \theta_i^{x_i-1/2}}{ \prod_{i=1}^m \Gamma(x_i + 1/2) } ,
\end{equation}
where [\cite{Mathematica:9}]
\begin{equation}
 Z(\bx|m) = \int_{\btheta\in\Theta}\hspace{-4mm} {\rm{Mu}}(\bx|m,\btheta)\pi_J(\btheta|m)\, {\rmd}\btheta =
 \frac{n! \Gamma(m/2) \prod_{i=1}^m \Gamma(x_i + 1/2)} {\sqrt{\pi^m} \Gamma(n+m/2)\prod_{i=1}^m x_i!}
\end{equation}
is the marginal likelihood and $\Theta$ is the $m$-dimensional simplex $\sum_{i=1}^m \theta_i = 1$. The posterior mean of $\theta_i$ is [\cite{Bernardo:1989}]
\begin{equation}\label{multi_post_mean1}
 {\rm E}(\theta_i|\bx,m) = \frac{x_i+1/2}{n+m/2} .
\end{equation}

Since (\ref{multi_post_mean1}) depends on the cell number, a paradox arises when the counts are explained by models including additional cells, where no event is observed. In particular, (\ref{multi_post_mean1}) depends on the number of void-cells and tends to zero as $m$ tends to the infinity.

\subsubsection{Proposed solution.}
\begin{figure}\centering
\includegraphics[width=7.5cm]{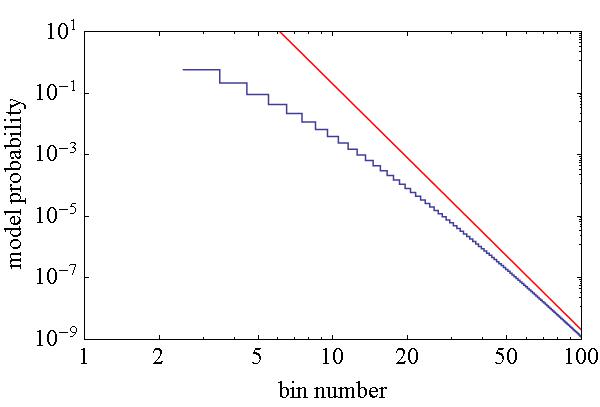}
\caption{Probability that the multinomial distribution $\rm{Mu}(x_1,x_2,\, ...\, x_m|\theta_1,\theta_2,\, ... \,\theta_m)$ for $m=3, 4,\, ...\, 100$ explains the data set $\bx=\{x_1=2,x_2=1,x_3=5,x_4=0,\, ...\, x_m=0\}$, where $n=8$. The red line is the $1/m^8$ asymptote.} \label{multi_fig}
\end{figure}

If the number of cells is unknown, the model uncertainty must be included into the analysis, as the conditioning over $m$ indicates. Therefore, (\ref{multi_post_mean1}) must be averaged over the models, the average being weighed by the model probabilities
\begin{equation}
 \Prob(m|\bx) = \frac{Z(\bx|m)}{\sum_{l=m_1}^{m_2} Z(\bx|l)} ,
\end{equation}
where $m_1$ is the number of non-null elements of $\bx$, $m_2$ is the upper bound to the cell number, and the models are assumed mutually exclusive and equiprobable. The asymptotic behaviour of $\Prob(m|\bx)$ is $1/m^n$; hence, models having increasing number of voids cells are less and less probable and contribute less and less to the posterior mean. An example is shown in Fig.\ \ref{multi_fig}.

\subsection{Stein paradox}\label{Stein}
When estimating the mean power of a number of signal from Gaussian measurements of their amplitudes, uniform priors of the unknown amplitudes lead to problematic posterior power distribution and expectation. In particular, as the number of signals tends to the infinity, the expected posterior power is inconsistent [\cite{Stein:1959,Bernardo:1979,Bernardo:2005,Attivissimo:2012,Samworth:2012,Carobbi:2014,Berger:2015}].

\subsubsection{Problem statement.}
In the simplest form of the Stein paradox, $x_i$ are realizations of independent normal variables with unknown means $\mu_i$ and known variance $\sigma^2$, for $i = 1, 2, ..., m$. To keep the algebra simple, without loss of generality, we use units where $\sigma=1$; when necessary for the sake of clarity, we will write explicitly quantity ratios {\it vs.} $\sigma$. The Jeffreys' prior for each $\mu_i$ is a constant, resulting in independent Gaussian posteriors, $N(\mu_i|x_i,\sigma=1)$, for each individual mean.

Let us consider $\mu_i^2$ -- from a Bayesian viewpoint, they are independent non-central $\chi^2_1$ variables having one degree of freedom, mean $1 + x_i^2$, and variance $2(1+2 x_i^2)$ -- and the $\theta^2=\big( \sum_{i=1}^m \mu_i^2 \big)/m$ measurand -- from a Bayesian viewpoint, it is a non-central $\chi^2_m$ variable having $m$ degrees of freedom. The Bayesian posterior mean and variance of $\theta^2$ are
\begin{subequations}\begin{equation}\label{theta_mean}
 \overline{\theta^2} = {\rm E}(\theta^2|\bx,\sigma=1,\mM_\infty) = 1 + \overline{x^2} ,
\end{equation}
where $\overline{x^2} = \sum_{i=1}^m x_i^2/m$, and
\begin{equation}\label{theta_var}
 \sigma^2_{\theta^2} = {\rm Var}(\theta^2|\bx,\sigma=1,\mM_\infty) = \frac{2}{m} \left( 1 + 2\overline{x^2} \right) ,
\end{equation}\end{subequations}
where $\bx$ is the $\{x_1, x_2,\, ...\, x_m\}$ list and $\mM_\infty$ is the model where the $\mu_i$ domain is $\mathds{R}$.

From a frequentist viewpoint, since $\sum_{i=1}^m x_i^2$ is a non-noncentral $\chi^2_m$ variable having $m$ degrees of freedom, $\overline{\theta^2}$ is a biased estimator of $\theta^2$. In fact,
\begin{equation}\label{frequentist}
 {\rm E}(\overline{\theta^2}|\bmu, \sigma=1) = 2 + \theta^2
\end{equation}
where $\bmu$ is the $\{\mu_1, \mu_2,\, ...\, \mu_m\}$ list. The frequentist variance of $\overline{\theta^2}$ is given by the same (\ref{theta_var}). As $m$ tends to infinity a worst situation occurs: (\ref{theta_mean}) and (\ref{theta_var}) predict that $\theta^2$ is certainly equal to $\overline{x^2} + 1$, but, at the same time, (\ref{frequentist}) and (\ref{theta_var}) predict that $\theta^2$ is certainly equal to $\overline{x^2}-1$.

\subsubsection{Proposed solution.}
To explain the paradox, we observe that it occurs only if the $\mu_i^2/m$ series converges. Otherwise, $\theta^2\rightarrow\infty$ and (\ref{theta_mean}) and (\ref{frequentist}) give the same result for all practical purposes. The improper prior $\pi_J(\mu_i)\propto$ const.\ encodes that $|\mu_i|$ is greater that any positive number. This information is irrelevant to the $\mu_i$ posterior -- besides, the odds on $\mu_i$ positive or negative are the same; but, it is not so for the $\mu_i^2$ posterior. If $|\mu_i| < \infty$, the $\mu_i$ domain must be bounded. Furthermore, if $\theta^2$ exists, these domains must be bounded also when $m\rightarrow\infty$.

A statistical model encoding this information is $\mM_{ba} =\{N(\bx|\bmu,\sigma=1) \,:\, \bmu\in [b-\sqrt{3}a,b+\sqrt{3}a]^m \}$, where $\bx$ is the $\{x_1, x_2,\, ...\, x_m\}$ list. In this way the data model gets a boundary and different hyper-parameters correspond to different models. The Jeffreys distribution of $\mu_i$ is the gate function $\pi_J(\mu_i|a,b) = U(\mu_i;b-\sqrt{3}a,b+\sqrt{3}a)$, which is zero outside the $[b-\sqrt{3}a,b+\sqrt{3}a]$ interval and $1/(2a)$ inside, the $a, b$ hyper-parameters being unknown. To infer (\ref{theta_mean}) and (\ref{theta_var}), the $a$ parameter was assumed large enough to identify the data model with $\mM_\infty$. However, this model is not necessarily supported by the data; therefore, model selection and averaging are necessary to take this missing information into account.

To make analytical computations possible, we substitute a Gaussian density for the $U(\mu_i;b-\sqrt{3}a,b+\sqrt{3}a)$ prior. We have not yet verified that the inferences made by using a Gaussian prior are not different from those made by using a gate prior. We do not expect qualitative differences, but, if so, it would be interesting to understand why. Hence, in the case of $m$ observations $\{x_i\} \sim$ $\prod_{i=1}^m N(x_i|\mu_i,\sigma$ $=1)$, the Gaussian approximation of the Jeffreys pre-data distribution of $\bmu$ is
\begin{equation}
 \pi_J(\bmu|a,b) = \prod_{i=1}^m N(\mu_i|b,a) ,
\end{equation}
which results in independent and identically distributed $\mu_i$ having marginal likelihood
\begin{eqnarray}\label{mar-lik1}\nonumber
 Z(x_i|b,a) &= &\int_{-\infty}^{+\infty}\!\! N(x_i|\mu_i,\sigma=1) N(\mu_i|b,a)\, \rmd\mu_i \\
 &= &\frac{ 1 } {\sqrt{2\pi(1+a^2)}} \exp\left[ -\frac{(x_i-b)^2}{2(1+a^2)} \right]
\end{eqnarray}
and normal posterior
\begin{subequations}\begin{equation}\label{Npost1}
 p(\mu_i|x_i,\sigma=1,b,a) = N(\mu_i|\overline{\mu_i}, \sigma_\mu) ,
\end{equation}
having
\begin{equation}\label{m1}
 \overline{\mu_i} = {\rm E}(\mu_i|x_i,\sigma=1,b,a) = \frac{a^2x_i+b}{1+a^2}
\end{equation}
mean and
\begin{equation}\label{smu}
 \sigma_\mu^2 = {\rm Var}(\mu_i|\sigma=1,b,a) = \frac{a^2}{1+a^2}
\end{equation}\end{subequations}
variance. The equation (\ref{mar-lik1}) benefits of the factorization
\begin{equation}
 N(x_i|\mu_i,\sigma) N(\mu_i|b,a) = N(x_i|b, \sqrt{a^2+\sigma^2})
 N\left( \mu_i\bigg| \frac{ax_i^2+b\sigma^2}{a^2+\sigma^2}, \frac{a\sigma}{\sqrt{a^2+\sigma^2}} \right),
\end{equation}
which solves the integration.

\begin{figure}\centering
\includegraphics[width=6.3cm]{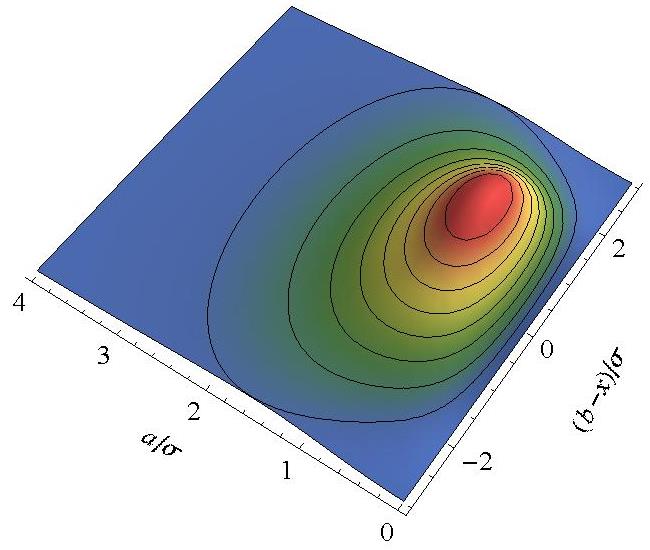}
\includegraphics[width=6.3cm]{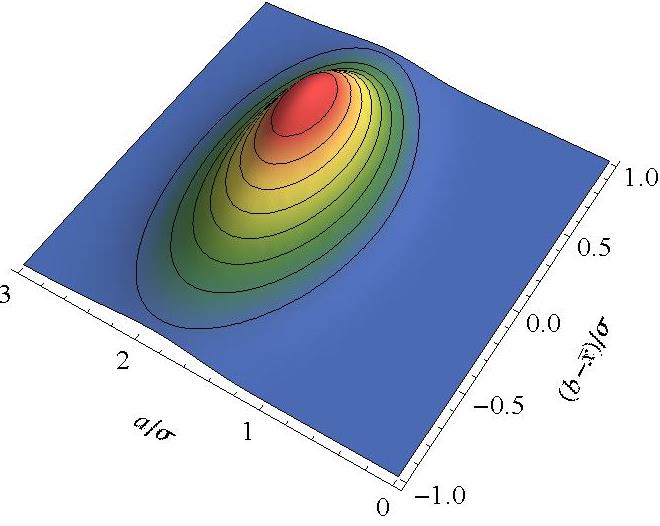}
\caption{Posterior probability densities of the model $\mM_{ba}$ given the $\{x_1,x_2,\, ...\, x_m\}$ Gaussian data. Left: $m=1$. Right: $m=20$ and $s_x/\sigma = 2$} \label{stein3-fig}
\end{figure}

By applying the Jeffreys rule to the marginal likelihood,
\begin{equation}
 Z(\bx|b,a) = \prod_{i=1}^m Z(x_i|b,a) = \frac{\exp\left\{ -\displaystyle\frac{m\big[s_x^2+(\overline{x}-b)^2\big]}{2(1+a^2)} \right\}}
 {\sqrt{(2\pi)^m(1+a^2)^m}}  ,
\end{equation}
which is the probability density of the data given the model and where $\overline{x}=\sum_{i=1}^m x_i/m$ and $s_x^2=\sum_{i=1}^m (x_i-\overline{x})^2/m$ are the sample mean and variance, we obtain the hyper-parameter prior
\begin{equation}
 \pi_J(b,a) \propto \frac{a}{\sqrt{(1+a^2)^3}} .
\end{equation}
Hence, the odds on $\mM_{ba}$ explaining the data are [\cite{Mathematica:9}]
\begin{eqnarray}\label{mp}
 p(b,a|\overline{x},s_x^2) &= &\frac{Z(\bx|b,a)\pi_J(b,a)}
 {\displaystyle\int_0^\infty\!\! \int_{-\infty}^{+\infty}\!\! Z(\bx|b,a)\pi_J(b,a)\,\rmd b\,\rmd a} \\
 &= &\frac{ \sqrt{2m(m s_x^2/2)^m}\, a } { \sqrt{\pi(1+a^2)^{m+3}}\, \Gamma(m/2,0,ms_x^2/2) }
 \exp \left\{ -\frac{m\big[ s_x^2+(b-\overline{x})^2 \big]}{2(1+a^2)} \right\} , \nonumber
\end{eqnarray}
where $\Gamma(a,z_1,z_2)$ is the generalized incomplete gamma function. Figure \ref{stein3-fig} shows the probability density (\ref{mp}) when $m=1$ (left) and $m=20$ and $s_x/\sigma = 2$ (right). It is worth noting that $p(b,a|\overline{x},s_x^2)$ depends only on the sample mean and variance and that its maximum occurs at $b=\overline{x}$.

\begin{figure}\centering
\includegraphics[width=7.5cm]{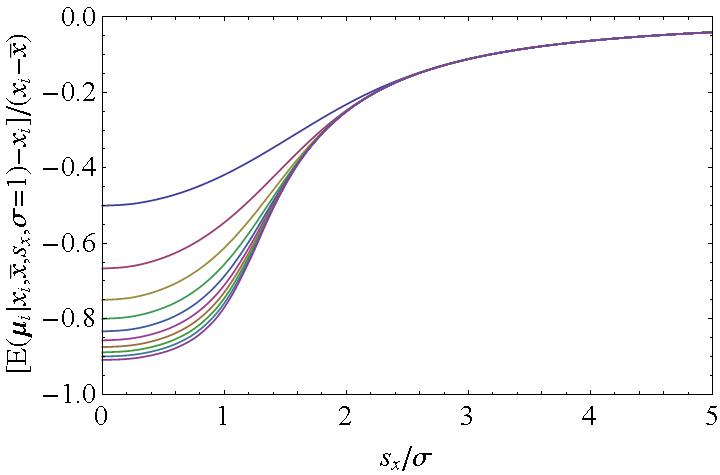}
\caption{Model-averaged mean of $\mu_i$ {\it vs.} the sample standard-deviation $s_x$. Ten cases were considered; from $m=2$ (top line) to $m=20$ (bottom line), in $m=2$ steps.} \label{stein4-fig}
\end{figure}

Let us discuss the $m=1$ case,
\begin{equation}\label{mp1D}
 p(b,a|x) = \frac{a}{\sqrt{2\pi}(1+a^2)^2} \exp \left[- \frac{(b-x)^2}{2(1+a^2)} \right] ,
\end{equation}
in some details. Firstly, we observe that not all the $\mu$ domains are equally supported by the $x$ datum, in particular the $\mathds{R}$ domain is excluded; the model most supported by the data, i.e., the mode of (\ref{mp1D}), is $\mM(x,1/\sqrt{3})$. In the second place, after averaging (\ref{Npost1}) over (\ref{mp1D}), the posterior distribution of $\mu$ and $\mu^2$ are
\begin{subequations}\begin{equation}
 p(\mu|x,\sigma=1) = \int_0^\infty\!\! \int_{-\infty}^{+\infty}\!\!  N(\mu|\overline{\mu},\sigma_\mu) p(b,a|x) \,\rmd b\,\rmd a
 = N(\mu|x,\sigma=1) ,
\end{equation}
and the non-central $\chi_1^2$ distribution,
\begin{equation}
 p(\mu^2|x,\sigma=1) = \int_{-\infty}^{+\infty}\!\! \delta(\mu^2-\tau^2) N(\tau|x,\sigma=1)\,\rmd\tau = \chi_1^2(\mu^2|x^2) ,
\end{equation}\end{subequations}
having one degree of freedom and non-centrality parameter $x^2$. This is an important and non-trivial result, which demonstrates that the hierarchical model $\mM_{ba}$ is consistent with the one-level model that uses the improper prior $\pi(\mu|\mM_\infty)\propto 1$.  Therefore, there is no hyper-prior effect on the posterior distributions of $\mu$ and $\mu^2$, whose expected values are ${\rm E}(\mu|x,\sigma=1)=x$ and ${\rm E}(\mu^2|x,\sigma=1)=1+x^2$. However, the $\mu^2$ expectation conditioned to the mode of the (\ref{mp1D}) distribution -- that is, to the $\mM(x,1/\sqrt{3})$ model most supported by the data -- decreases to
\begin{equation}
 {\rm E}(\mu^2|x,\sigma=1,b=x,a=1/\sqrt{3}) = 1/4 + x^2 .
\end{equation}

After averaging the means (\ref{m1}) over the models, we obtain [\cite{Mathematica:9}]
\begin{eqnarray}
 {\rm E}(\mu_i|x_i,\overline{x},s_x,\sigma=1)
 &= &\int_0^\infty\!\! \int_{-\infty}^{+\infty}\!\! {\rm E}(\mu_i|x_i,\sigma=1,b,a) p(b,a|\overline{x},s_x^2)\,\rmd b\,\rmd a \\ \nonumber
 &= &x_i + \frac{ 2(x_i-\overline{x})\big[ (m+2)\Gamma(m/2,0,ms_x^2/2) - 2\Gamma(m/2+2) \big] } {m(m+2)s_x^2 \Gamma(m/2,0,ms_x^2/2)} .
\end{eqnarray}
Figure \ref{stein4-fig} shows the difference between the model-averaged mean and $x_i$. When the sample variance $s_x^2/\sigma^2$ is small, ${\rm E}(\mu_i|x_i,\overline{x},s_x,\sigma=1) \rightarrow x_i-m(x_i-\overline{x})/(m+2)$ and the model average shrinks towards $\overline{x}$; more are the measurands, the stronger the shrink. Actually, when $m\rightarrow\infty$, data consistency requires that the lower bound of the sample variance is $s_x^2/\sigma^2=1$; in this case the model average converges to the sample mean. When the sample variance is large, ${\rm E}(\mu_i|x_i,\overline{x},s_x,\sigma=1) \rightarrow x_i$ and the model average supports the $x_i$ datum. This is consistent with a large (small) sample variance supporting the hypothesis that the measurands are different (the same).

\begin{figure}\centering
\includegraphics[width=7.5cm]{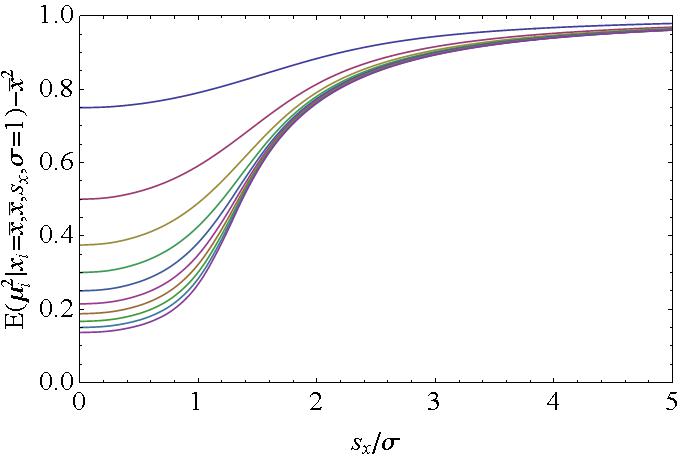}
\caption{Model-averaged mean of $\mu_i^2$ {\it vs.} the sample standard-deviation $s_x$, when $x_i=\overline{x}$. Ten cases were considered; from $m=2$ (top line) to $m=20$ (bottom line), in $m=2$ steps.} \label{stein5-fig}
\end{figure}

The $(\mu_i/\sigma_\mu)^2$ measurands are independent and identical non-central $\chi_1^2$ variables having one degree of freedom and non-centrality parameter $\lambda_i^2=(\overline{\mu_i}/\sigma_\mu)^2$. Hence,
\begin{equation}
 p(\mu_i^2|x_i,\sigma=1,b,a) = \frac{\chi_1^2(\mu^2_i/\sigma_\mu^2|\lambda_i^2)}{\sigma_\mu^2} ,
\end{equation}
where $\overline{\mu_i}$ and $\sigma_\mu^2$ are given by (\ref{m1}) and (\ref{smu}). The expected $\mu_i^2$ value is
\begin{equation}
 {\rm E}(\mu_i^2|x_i,\sigma=1,b,a) = \sigma_\mu^2 + \overline{\mu_i}^2
 = \frac{b^2 + a^2 (1 + 2 b x_i) + a^4 (1 + x_i^2)}{(1+a^2)^2}
\end{equation}
and, after averaging over the models, we obtain [\cite{Mathematica:9}]
\begin{eqnarray}\label{mu2}
 {\rm E}(\mu_i^2|x_i,\overline{x},s_x,\sigma=1)
 = \int_0^\infty\!\! \int_{-\infty}^{+\infty}\!\! {\rm E}(\mu_i^2|x_i,\sigma=1,b,a) p(b,a|\overline{x},s_x^2)\,\rmd b\,\rmd a \nonumber \\
 = x_i^2 + 1 - \frac{A\exp(-ms_x^2/2) + B\big[ 2\Gamma(m/2+2) - (m+2)\Gamma(m/2+1,ms_x^2/2) \big]}
 { (m+2) m^2 s_x^4 \Gamma(m/2,0,ms_x^2/2)} ,
\end{eqnarray}
where
\begin{equation}
 A = \frac{(m+2)\sqrt{(m s_x^2)^{m+2}} (x_i - \overline{x})^2}{\sqrt{2^{m-2}}}
\end{equation}
and
\begin{equation}
 B = 2s_x^2 \big[1 - m - 2 m x_i (x_i - \overline{x}) \big] + 2(m+2) (x_i - \overline{x})^2 .
\end{equation}

In addition to the sample variance, the model average (\ref{mu2}) depends on both $x_i$ and the sample mean; this is a consequence of the prior information that all $\mu_i$ belong to the same interval. When the sample variance tends to zero and infinity, the model average converges to [\cite{Mathematica:9}]
\begin{equation}
 \lim_{s_x \rightarrow 0} {\rm E}(\mu_i^2|x_i,\overline{x}=x_i,s_x,\sigma=1) = x_i^2 + 1 - \frac{m-1}{m+2} ,
\end{equation}
where $\overline{x} \rightarrow x_i$ because of consistency, and $x_i^2+1$, respectively. When $m \rightarrow \infty$, the sample variance is bounded by $s_x^2/\sigma^2 > 1$ and [\cite{Mathematica:9}]
\begin{equation}
 \lim_{m \rightarrow \infty} {\rm E}(\mu_i^2|x_i,\overline{x},s_x=1,\sigma=1) = \overline{x}^2 .
\end{equation}
These results are consistent with a large variance indicating different measurands -- hence, the model average is $x_i^2+1$ -- and a unit variance indicating the same measurand -- hence, the model average is $\overline{x}^2$. Figure \ref{stein5-fig} shows the difference between the model average and $\overline{x}^2$, when $x_i=\overline{x}$. When $s_x^2/\sigma^2$  is small, ${\rm E}(\mu_i^2|x_i,\overline{x},s_x,\sigma=1) \rightarrow \overline{x}^2$; most are the measurands, the nearest is the average. When the sample variance is big, or there is only one measurand, the model average converges to $x_i^2+1$.

\begin{figure}\centering
\includegraphics[width=7.5cm]{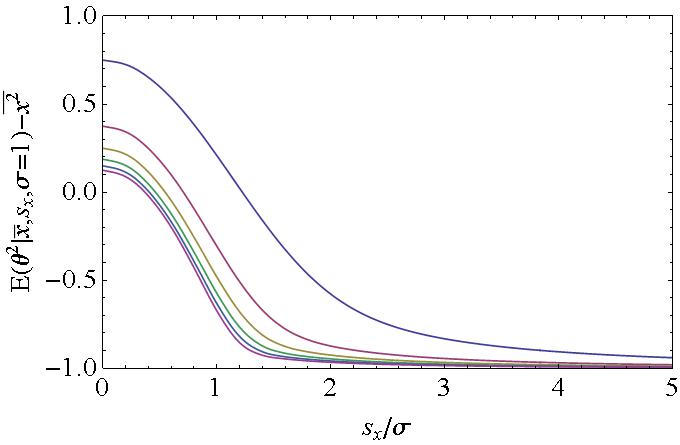}
\caption{Model-averaged expectation of $\theta^2$. Five cases were considered; from $m=2$ (top line) to $m=22$ (bottom line), in $m=4$ steps.} \label{stein6-fig}
\end{figure}

Eventually, let us turn the attention to $\theta^2$. The expected value is
\begin{equation}
 {\rm E}(\theta^2|\overline{x},s_x,\sigma=1,b,a) = \frac{\displaystyle\sum_{i=1}^m (\sigma_\mu^2 + \overline{\mu_i}^2)}{m}
 = \frac{ b^2 + a^2(1+2b\overline{x}) + a^4(1+s_x^2+\overline{x}^2) } {(1+a^2)^2}
\end{equation}
and, after averaging over the odds on the data models [\cite{Mathematica:9}],
\begin{eqnarray}
 {\rm E}(\theta^2|\overline{x^2},s_x,\sigma=1)
 &= &\int_0^\infty\!\! \int_{-\infty}^{+\infty}\!\! {\rm E}(\theta^2|\overline{x},s_x,\sigma=1,b,a) p(b,a|\overline{x},s_x^2)\,\rmd b\,\rmd a \nonumber \\
 &= &1 + \overline{x^2} + 2A/B ,
\end{eqnarray}
where
\begin{subequations}\begin{eqnarray}
 A &= &(3-2ms_x^2)\Gamma(m/2+1) -2\Gamma(m/2+2,ms_x^2/2) \\ \nonumber
   &  &- (1-m-2ms_x^2)\Gamma(m/2+1,ms_x^2/2)
\end{eqnarray}
and
\begin{equation}
 B = m^2 s_x^2 \Gamma(m/2,0,ms_x^2/2) .
\end{equation}\end{subequations}

As shown in Fig.\ \ref{stein6-fig}, the Stein paradox is solved by observing that, when $s_x^2/\sigma^2$ tends to zero -- that is, when there is a single measurand, the model average converges to the frequentist estimate $\overline{x^2}$; most are the measurands, the nearest are the estimates. In the limit of a big sample-variance -- that is, when there are many different measurands, the model average converges to $\overline{x^2} - 1$, which is the frequentist unbiased estimate. It is non-obvious and remarkable that, when $m\rightarrow\infty$ and, consistently, $s_x^2/\sigma^2 > 1$, the model average is always  $\overline{x^2} - 1$.

\subsection{Neyman-Scott paradox}
The problem is to estimate the common variance of independent Gaussian observations of different quantities, where the quantity values are not of interest. In the simplest case, this is a fixed effect model with two observations on each quantity. As the number of observed quantities grows without bound, the Jeffreys prior leads to an inconsistent expectation of the common variance [\cite{Neyman-Scott:1948}].

\subsubsection{Problem statement.}
Let $\{x_1, x_2\}_i$ be $m$ independent pairs of independent Gaussian variables having different means $\mu_i$ and common variance $\zeta=\sigma^2$, for $i = 1,\, ...\, m$, all parameters being unknown. By changing the data from $\{x_1, x_2\}_i$ to the sample means $\overline{x_i}=(x_{1i}+x_{2i})/2$ and pooled variance $s^2=\sum_{i=1}^m s_i^2/m$, where $s_i^2=(x_{1i}-x_{2i})^2/2$, the sampling distribution is
\begin{equation}\label{NS_Lik}
 p(\overline{\bx},s^2|\bmu,\zeta) = \frac{2m}{\zeta} \chi_m^2(2ms^2/\zeta) N_m(\overline{\bx}|\bmu,\zeta\mathds{I}_m/2) ,
\end{equation}
where $\overline{\bx} = [\overline{x_1}, \overline{x_2},\, ...\,\overline{x_m}]^{\rm T}$, $\chi_m^2(z)$ is a chi-square distribution having $m$ degrees of freedom, and $N_m(\overline{\bx}|\bmu,\zeta\mathds{I}_m/2)$ is a $m$-variate normal distribution having mean $\bmu=[\mu_1,\mu_2,\, ...\,\mu_m]^{\rm T}$ and $\zeta\mathds{I}_m/2$ variance-covariance matrix.

The Jeffreys prior of the $\{\bmu, \zeta\}$ parameters is
\begin{equation}\label{Scott-prior}
 \pi_J(\bmu,\zeta|\zeta_0=0) \propto 1/\sqrt{\zeta^{m+2}} ,
\end{equation}
where $\mM_0 = \{p(\overline{\bx},s^2|\bmu,\zeta) \,:\, (\bmu,\zeta) \in \mathds{R}^m\times \mathds{R}^+\}$ is the data model and $\zeta > \zeta_0 = 0$. The reason of the explicit introduction of $\zeta_0=0$ hyper-parameter will be clarified in the next section. The prior (\ref{Scott-prior}) results in the marginal likelihood [\cite{Mathematica:9}]
\begin{eqnarray}\label{Scott-Z1}\nonumber
 Z(\overline{\bx},s^2|\zeta_0=0)
 &= &\int_0^\infty\!\! \int_{-\infty}^{+\infty} \!\! p(\overline{\bx},s^2|\bmu,\zeta) \pi_J(\bmu,\zeta|\zeta_0=0) \,\rmd\bmu\,\rmd\zeta \\
 &\propto &\frac{\Gamma(m+1)}{\sqrt{(ms^2)^{m+2}}\, \Gamma(m/2)} ,
\end{eqnarray}
where the specification $\zeta_0=0$ highlights the conditioning over the data-explaining model, and posterior distribution
\begin{equation}\label{Scott-joint-post}
 p(\bmu,\zeta|\overline{\bx},s^2,\zeta_0=0) = \frac{m(ms^2)^m}{\sqrt{\pi^m}\, \Gamma(m+1)}
 \frac{\exp\left( -\displaystyle\frac{ms^2+|\bmu-\overline{\bx}|^2}{\zeta} \right)}{\sqrt{\zeta^{3m+2}}} .
\end{equation}
After the means $\bmu$ are integrated out, the posterior density and expected value of $\zeta$ are
\begin{equation}\label{Scott-post1}
 p(\zeta|\overline{\bx},s^2,\zeta_0=0) = \int_{-\infty}^{+\infty}\!\! p(\bmu,\zeta|\overline{\bx},s^2,\zeta_0=0)\, \rmd\bmu
 = \frac{(m s^2)^2 \exp(-ms^2/\zeta)}{\zeta^{m+1}\Gamma(m)}
\end{equation}
and
\begin{subequations}\begin{equation}\label{Scott-mean}
 \overline{\zeta} = {\rm E}(\zeta|\overline{\bx},s^2,\zeta_0=0)
 = \int_0^\infty\!\! \zeta p(\zeta|\overline{\bx},s^2,\zeta_0=0)\, \rmd\zeta = \frac{ms^2}{m-1} ,
\end{equation}
where $m\ge 2$, with a variance equal to
\begin{equation}\label{Scott-uncertainty}
 {\rm Var}(\zeta|\overline{\bx},s^2,\zeta_0=0) = \int_0^\infty\!\! \zeta^2 p(\zeta|\overline{\bx},s^2,\zeta_0=0)\, \rmd\zeta - \overline{\zeta}^2
 = \frac{\overline{\zeta}^2}{m-2} ,
\end{equation}\end{subequations}
where $m\ge 3$.

Since $2ms^2/\zeta$ is a $\chi^2_m$ variable having $m$ degrees of freedom, $s^2$ is a frequentist biased estimator of $\zeta$. In fact,
\begin{subequations}\begin{equation}\label{frequentist-mean}
 \overline{s^2} = {\rm E}(s^2|\zeta) = \frac{2m}{\zeta} \int_0^\infty s^2 \chi_m^2(2ms^2/\zeta)\, \rmd s^2 = \zeta/2
\end{equation}
and
\begin{equation}\label{frequentist-uncertainty}
 {\rm Var}(s^2|\zeta) = \frac{2m}{\zeta}
 \int_0^\infty  s^4 \chi_m^2(2ms^2/\zeta)\, \rmd s^2 - (\overline{s^2})^2 = \zeta^2/m .
\end{equation}\end{subequations}
Also in this case a paradox occurs. As $m$ tends to the infinity, (\ref{Scott-mean}) and (\ref{Scott-uncertainty}) predict that $\zeta$ is certainly equal to $s^2$, but, (\ref{frequentist-mean}) and (\ref{frequentist-uncertainty}) predict that $\zeta$ is certainly equal to $2s^2$.

\subsubsection{Proposed solution.}
To explain the paradox, we observe that (\ref{Scott-prior}) gives for certain that $\zeta$ is less than any positive number, no matter how small it is. Therefore, (\ref{Scott-prior}) encodes that $\zeta$ is null, but we observe a non-zero pooled variance. To avoid this contradiction, the $\zeta$'s domain must be limited to $[\zeta_0,\infty[$, the hyper-parameter $\zeta_0=\sigma_0^2 > 0$ being unknown. Hence, the Jeffryes' distribution
\begin{equation}
 \pi_J(\bmu,\zeta|\zeta_0) = \frac{m}{2}\sqrt{\frac{\zeta_0^m}{\zeta^{m+2}}}\, \vartheta(\zeta-\zeta_0) ,
\end{equation}
where the Heaviside function $\vartheta(z)$ is null for $z<0$ and one otherwise, substitutes for (\ref{Scott-prior}). This results in the marginal likelihood [\cite{Mathematica:9}]
\begin{eqnarray}\label{Scott-evidence}\nonumber
 Z(\overline{\bx},s^2|\zeta_0)
 &= &\int_{\zeta_0}^\infty\!\! \int_{-\infty}^{+\infty} \!\! p(\overline{\bx},s^2|\bmu,\zeta) \pi_J(\bmu,\zeta|\zeta_0) \,\rmd\bmu\,\rmd\zeta \\
 &= &\frac{ m^2\sqrt{\zeta_0^m}\, \Gamma(m, 0, ms^2/\zeta_0) } {4\sqrt{m^m}\, s^{m+2} \Gamma(m/2+1)}
\end{eqnarray}
and posterior distribution
\begin{equation}\label{Scott-post}
 p(\bmu,\zeta|\overline{\bx},s^2,\zeta_0) = \frac{(ms^2)^m}{\sqrt{\pi^m} \Gamma(m, 0, ms^2/\zeta_0) }
 \frac{\exp\left( -\displaystyle\frac{ms^2+|\bmu-\overline{\bx}|^2}{\zeta} \right)}{\sqrt{\zeta^{3m+2}}} ,
\end{equation}
where $\zeta>\zeta_0$.

After integrating out the means $\bmu$, the posterior density and expected value of $\zeta$ are
\begin{equation}
 p(\zeta|\overline{\bx},s^2,\zeta_0) = \int_{-\infty}^{+\infty}\!\! p(\bmu,\zeta|\overline{\bx},s^2,\zeta_0)\, \rmd\bmu
 = \frac{(ms^2)^m \exp(-ms^2/\zeta) }{ \zeta^{m+1}\Gamma(m, 0, ms^2/\zeta_0) } ,
\end{equation}
where $\zeta>\zeta_0$, and
\begin{equation}\label{z-expectation}
 {\rm E}(\zeta|\overline{\bx},s^2,\zeta_0) = \int_{\zeta_0}^\infty\!\! \zeta\, p(\zeta|\overline{\bx},s^2,\zeta_0)\, \rmd\zeta
 = \frac{ms^2 \Gamma(m-1, 0, ms^2/\zeta_0) }{ \Gamma(m, 0, ms^2/\zeta_0) } .
\end{equation}

\begin{figure}\centering
\includegraphics[width=7.5cm]{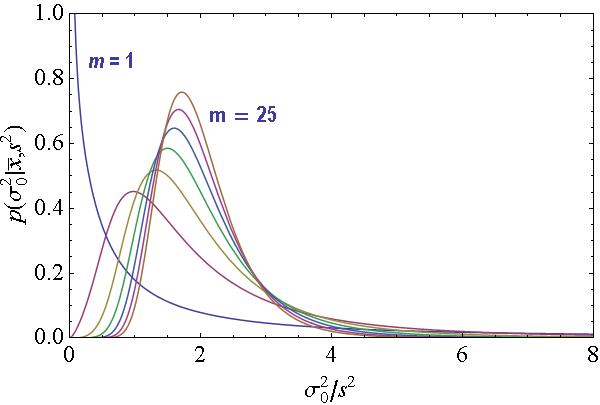}
\caption{Probability density of the data models (\ref{NS_Lik}), where $\zeta=\sigma^2 > \sigma^2_0$, {\it vs.} the $\sigma^2_0$ hyper-parameter. Different numbers of observed quantities, from $m=1$ to $m=25$, are considered, in $m=4$ steps.} \label{NS-fig}
\end{figure}

Different hyper-parameters, that is, different lower bound of the $\zeta$ coordinate, correspond to different manifolds and to different models. In order to average over them, we need the Bayes theorem to calculate the model probability -- consistently with the observed data -- and, in turn, we need the prior of $\zeta_0$. By applying the Jeffreys rule to (\ref{Scott-evidence}), which is the sampling distribution of $\overline{\bx}$ and $s^2$ given $\zeta_0$, we obtain
\begin{equation}\label{pz0}
 \pi_J(\zeta_0) \propto 1/\zeta_0 ,
\end{equation}
After combining (\ref{pz0}) with (\ref{Scott-evidence}), the posterior probability density of $\zeta_0$ is
\begin{equation}\label{Scott-model}
 p(\zeta_0|\overline{\bx},s^2) = \frac{ m^2\sqrt{\zeta_0^{m-2}}\, \Gamma(m, 0, ms^2/\zeta_0) }
 {4\sqrt{m^m}\, s^{m+2} \Gamma(m/2+1)} .
\end{equation}
The odds on $\zeta_0$ explaining the data are shown in Fig.\ \ref{NS-fig}. It is worth noting that, in the limit of many observed quantities, the lower bound of the model variance most supported by the data is twice the pooled variance.

The Neymann-Scott paradox is solved by observing that, after averaging (\ref{z-expectation}) over $\zeta_0$ with the (\ref{Scott-model}) weighs, the expected $\zeta$ value is [\cite{Mathematica:9}]
\begin{equation}\label{Scott-mean2}
 {\rm E}(\zeta|\overline{\bx},s^2) = \frac{2ms^2}{m-2} ,
\end{equation}
where $m\ge 3$. In the limit when $m\rightarrow\infty$, ${\rm E}(\zeta|\overline{\bx},s^2) = 2s^2$, which is the frequentist unbiased estimate.

\subsection{marginalization paradox}
The post-data distribution (\ref{Scott-post1}) of $\zeta=\sigma^2$ depends only on the pooled variance $s^2$. Since $s^2$ and $\overline{\bx}$ are independent chi-square and normal variables, respectively, it might seem that the $\overline{\bx}$ data are irrelevant and can be omitted [\cite{Dawid:1973,Bernardo:1979,Kass:1996}]. But, it is not so.

To investigate the contribution of $\overline{\bx}$ to inferences about $\zeta$, let us discard it and start from the $ \mX = \{ p(s^2|\zeta,m) \,:\, \zeta \in \mathds{R}^+\}$ model, where the sampling distribution is
\begin{equation}
 p(s^2|\zeta,m) = \frac{2m}{\zeta} \chi_m^2(2ms^2/\zeta) .
\end{equation}
The Jeffreys prior of $\zeta$ is $\pi_J(\zeta|\mX)\propto 1/\zeta$. This results in the marginal likelihood
\begin{equation}
 Z(s^2|m,\mX) = \int_0^\infty p(s^2|\zeta,m,\mX) \pi_J(\zeta|\mX)\, \rmd\zeta \propto 1/s^2 ,
\end{equation}
posterior-distribution
\begin{equation}\label{mpar-post}
 p(\zeta|s^2,m,\mX) = \frac{\sqrt{(ms^2)^m} \exp(-ms^2/\zeta)}{\sqrt{\zeta^{m+2}} \Gamma(m/2)} ,
\end{equation}
and expectation
\begin{equation}\label{mpar-mean}
 \overline{\zeta} = {\rm E}(\zeta|s^2,m,\mX) = \int_0^\infty \zeta p(\zeta|s^2,m,\mX)\, \rmd\zeta = \frac{2ms^2}{m-2}
\end{equation}
that are different from (\ref{Scott-post1}) and (\ref{Scott-mean}).

To explain the paradox, we observe that -- as shown by the joint posterior distribution (\ref{Scott-joint-post}) -- $\bmu$ and $\zeta$ are not independent [\cite{Jaynes:1980}]. Therefore, $\bmu$ is relevant and its posterior distribution affect (\ref{Scott-post1}). The posterior distribution (\ref{mpar-post}) differs from (\ref{Scott-post1}) because the information delivered by $\overline{\bx}$ has been neglected. As a consequence the $\zeta$ variance conditioned to $\mX$,
\begin{equation}
 {\rm Var}(\zeta|s^2,m,\mX) = \int_0^\infty \zeta^2 p(\zeta|s^2,m,\mX)\, \rmd\zeta - \overline{\zeta}^2 = \frac{2\overline{\zeta}^2}{m-4} ,
\end{equation}
is bigger than (\ref{Scott-uncertainty}). It is also worth noting that (\ref{mpar-mean}) does not suffer from the (\ref{Scott-mean}) inconsistency and it is the same as (\ref{Scott-mean2}). Our conjectured explanation is as follows. Differently from (\ref{Scott-prior}), which is conditioned to $\mM_0$, the $\pi_J(\zeta|\mX)\propto 1/\zeta$ distribution assigns the same probability to any $\zeta$'s sub-domain including the zero or infinity and a null probability to any sub-domain including none of the two. Therefore, the biasing effect of (\ref{Scott-prior}) is removed.

\section{Conclusions}
In metrology, assessing the measurement uncertainty requires invariant metrics of the data models and covariant priors. In short,
\begin{itemize}
 \item the posterior probability density must be covariant for model re-pa\-ra\-me\-tri\-zation and, consequently, the prior density must be covariant;
 \item the Jeffreys priors, which are derived from the volume element of model manifold equipped with the information metric, are sensible choices;
 \item the posterior probability of a selected model must not be null; therefore, improper priors are not allowed;
 \item to avoid paradox and inconsistencies, the computation of the posterior probability may involve hierarchical models, model selection, and averaging.
\end{itemize}
These ideas were tested by the application to a number of key paradoxical problems. Non-uniquenesses were identified in the data-explaining models. In addition, parametric models having an infinite volume, corresponding to improper priors, were found to hide unacknowledged information. After recognizing these issues and taking them into account via hierarchical modelling and model averaging, inconsistencies and paradoxes disappeared.

\bibliographystyle{ba}
\bibliography{prior}

\begin{thebibliography}{40}
\newcommand{\enquote}[1]{``#1''}
\expandafter\ifx\csname natexlab\endcsname\relax\def\natexlab#1{#1}\fi
\expandafter\ifx\csname url\endcsname\relax
  \def\url#1{{\tt #1}}\fi
\expandafter\ifx\csname urlprefix\endcsname\relax\def\urlprefix{URL }\fi
\ifx\endbibitem\undefined \let\endbibitem\relax\fi

\bibitem[{Amari et~al.(2007)Amari, Nagaoka, and Harada}]{Amari:2007}
Amari, S., Nagaoka, H., and Harada, D. (2007).
\newblock {\em Methods of Information Geometry\/}.
\newblock Translations of Mathematical Monographs vol.\ 191. Oxford: Oxford
  University Press.
\endbibitem

\bibitem[{Arwini and Dodson(2008)}]{Arwini:2008}
Arwini, K. and Dodson, C. (2008).
\newblock {\em Information Geometry: Near Randomness and Near Independence\/}.
\newblock Lecture Notes in Mathematics 1953. Berlin Heidelberg: Springer.
\endbibitem

\bibitem[{Attivissimo et~al.(2012)Attivissimo, Giaquinto, and
  Savino}]{Attivissimo:2012}
Attivissimo, F., Giaquinto, N., and Savino, M. (2012).
\newblock \enquote{A Bayesian paradox and its impact on the GUM approach to
  uncertainty.}
\newblock {\em Measurement\/}, 45(9): 2194--2202.
\endbibitem

\bibitem[{Berger and Bernardo(1992{\natexlab{a}})}]{Berger:1992}
Berger, J.~O. and Bernardo, J.~M. (1992{\natexlab{a}}).
\newblock \enquote{On the development of reference priors.}
\newblock In Bernardo, J.~M., Berger, J.~O., Dawid, A.~P., and Smith, A. F.~M.
  (eds.), {\em Bayesian Statistics: Proceedings of the Fourth Valencia
  International Meeting\/}. Oxford: Clarendon Press.
\endbibitem

\bibitem[{Berger and Bernardo(1992{\natexlab{b}})}]{Berger:1992a}
--- (1992{\natexlab{b}}).
\newblock \enquote{Ordered group reference priors with application to the
  multinomial problem.}
\newblock {\em Biometrika\/}, 79: 25--37.
\endbibitem

\bibitem[{Berger et~al.(2009)Berger, Bernardo, and Sun}]{Berger:2009}
Berger, J.~O., Bernardo, J.~M., and Sun, D. (2009).
\newblock \enquote{The formal definition of reference priors.}
\newblock {\em Ann. Statist.\/}, 37(2): 905--938.
\endbibitem

\bibitem[{Berger et~al.(2015)Berger, Bernardo, and Sun}]{Berger:2015}
--- (2015).
\newblock \enquote{Overall Objective Priors.}
\newblock {\em Bayesian Anal.\/}, 10(1): 189--221.
\endbibitem

\bibitem[{Berger and Sun(2008)}]{Berger:2008}
Berger, J.~O. and Sun, D. (2008).
\newblock \enquote{Objective priors for the bivariate normal model.}
\newblock {\em Ann.\ Statist.\/}, 36(2): 963--982.
\endbibitem

\bibitem[{Bernardo(1979)}]{Bernardo:1979}
Bernardo, J.~M. (1979).
\newblock \enquote{Reference Posterior Distributions for Bayesian Inference.}
\newblock {\em Journal of the Royal Statistical Society. Series B
  (Methodological)\/}, 41(2): 113--147.
\endbibitem

\bibitem[{Bernardo(1989)}]{Bernardo:1989}
--- (1989).
\newblock \enquote{The Geometry of Asymptotic Inference: Comment: On
  Multivariate Jeffreys' Priors.}
\newblock {\em Statistical Science\/}, 4(3): 227--229.
\endbibitem

\bibitem[{Bernardo(2005)}]{Bernardo:2005}
--- (2005).
\newblock \enquote{Reference analysis.}
\newblock In K., D.~D. and R., R.~C. (eds.), {\em Handbook of Statistics 25\/}.
  Amsterdam: North Holland.
\endbibitem

\bibitem[{Bodnar et~al.(2015)Bodnar, Link, and Elster}]{Bodnar:2015}
Bodnar, O., Link, A., and Elster, C. (2015).
\newblock \enquote{Objective Bayesian Inference for a Generalized Marginal
  Random Effects Model.}
\newblock {\em Bayesian Anal.\/}, advance publication.
\endbibitem

\bibitem[{Carobbi(2014)}]{Carobbi:2014}
Carobbi, C. (2014).
\newblock \enquote{Bayesian inference on a squared quantity.}
\newblock {\em Measurement\/}, 48(1): 13--20.
\endbibitem

\bibitem[{Clyde and George(2004)}]{Clyde:2004}
Clyde, M. and George, E.~I. (2004).
\newblock \enquote{Model Uncertainty.}
\newblock {\em Statistical Science\/}, 19(1): 81--94.
\endbibitem

\bibitem[{Costa et~al.(2014)Costa, Santos, and Strapasson}]{Costa:2014}
Costa, S.~I., Santos, S.~A., and Strapasson, J.~E. (2014).
\newblock \enquote{Fisher information distance: A geometrical reading.}
\newblock {\em Discrete Applied Mathematics\/}, 197: 59--69.
\endbibitem

\bibitem[{D'Agostini(2003)}]{D'Agostini:2003}
D'Agostini, G. (2003).
\newblock {\em Bayesian Reasoning in Data Analysis: A Critical Introduction\/}.
\newblock Singapore: World Scientific.
\endbibitem

\bibitem[{Datta and Ghosh(1995)}]{Datta:1995}
Datta, G.~S. and Ghosh, M. (1995).
\newblock \enquote{Some Remarks on Noninformative Priors.}
\newblock {\em Journal of the American Statistical Association\/}, 90(432):
  1357--1363.
\endbibitem

\bibitem[{Datta and Ghosh(1996)}]{Datta:1996}
--- (1996).
\newblock \enquote{On the invariance of noninformative priors.}
\newblock {\em Ann. Statist.\/}, 24(1): 141--159.
\endbibitem

\bibitem[{Dawid et~al.(1973)Dawid, Stone, and Zidek}]{Dawid:1973}
Dawid, A.~P., Stone, M., and Zidek, J.~V. (1973).
\newblock \enquote{Marginalization Paradoxes in Bayesian and Structural
  Inference.}
\newblock {\em Journal of the Royal Statistical Society. Series B
  (Methodological)\/}, 35(2): pp. 189--233.
\endbibitem

\bibitem[{Dose(2007)}]{Dose:2007}
Dose, V. (2007).
\newblock \enquote{Bayesian estimate of the Newtonian constant of gravitation.}
\newblock {\em Measurement Science and Technology\/}, 18(1): 176--182.
\endbibitem

\bibitem[{Elster and Toman(2010)}]{Elster:2010}
Elster, C. and Toman, B. (2010).
\newblock \enquote{Analysis of key comparisons: estimating laboratories' biases
  by a fixed effects model using Bayesian model averaging.}
\newblock {\em Metrologia\/}, 47(3): 113--119.
\endbibitem

\bibitem[{Ghosh(2011)}]{Ghosh:2011}
Ghosh, M. (2011).
\newblock \enquote{Objective Priors: An Introduction for Frequentists.}
\newblock {\em Statist. Sci.\/}, 26(2): 187--202.
\endbibitem

\bibitem[{Jaynes and Bretthorst(2003)}]{Jaynes:2003}
Jaynes, E. and Bretthorst, G. (2003).
\newblock {\em Probability Theory: The Logic of Science\/}.
\newblock Cambridge: Cambridge University Press.
\endbibitem

\bibitem[{Jaynes(2012)}]{Jaynes:1980}
Jaynes, E.~T. (2012).
\newblock \enquote{Marginalization and Prior Probabilities.}
\newblock In Rosenkrantz, R.~D. (ed.), {\em E. T. Jaynes: Papers on
  Probability, Statistics and Statistical Physics\/}. Dordrecht: Springer.
\endbibitem

\bibitem[{Jeffreys(1946)}]{Jeffreys:1946}
Jeffreys, H. (1946).
\newblock \enquote{An Invariant Form for the Prior Probability in Estimation
  Problems.}
\newblock {\em Proceedings of the Royal Society of London A: Mathematical,
  Physical and Engineering Sciences\/}, 186: 453--461.
\endbibitem

\bibitem[{Jeffreys(1998)}]{Jeffreys:1998}
--- (1998).
\newblock {\em The Theory of Probability\/}.
\newblock Oxford: Oxford University Press.
\endbibitem

\bibitem[{Kass(1989)}]{Kass:1989}
Kass, R.~E. (1989).
\newblock \enquote{The Geometry of Asymptotic Inference.}
\newblock {\em Statist. Sci.\/}, 4(3): 188--219.
\endbibitem

\bibitem[{Kass and Wasserman(1996)}]{Kass:1996}
Kass, R.~E. and Wasserman, L. (1996).
\newblock \enquote{The Selection of Prior Distributions by Formal Rules.}
\newblock {\em Journal of the American Statistical Association\/}, 91(435):
  1343--1370.
\endbibitem

\bibitem[{MacKay(2003)}]{MacKay:2003}
MacKay, D. (2003).
\newblock {\em Information Theory, Inference and Learning Algorithms\/}.
\newblock Cambridge: Cambridge University Press.
\endbibitem

\bibitem[{Mana(2015)}]{Mana:2015}
Mana, G. (2015).
\newblock \enquote{Model uncertainty and reference value of the {P}lanck
  constant.}
\newblock {\em Measurement\/}, submitted.
\endbibitem

\bibitem[{Mana et~al.(2014)Mana, Giuliano~Albo, and Lago}]{Mana:2014}
Mana, G., Giuliano~Albo, P.~A., and Lago, S. (2014).
\newblock \enquote{Bayesian estimate of the degree of a polynomial given a
  noisy data sample.}
\newblock {\em Measurement\/}, 55: 564--570.
\endbibitem

\bibitem[{Mana et~al.(2012)Mana, Massa, and Predescu}]{Mana:2012}
Mana, G., Massa, E., and Predescu, M. (2012).
\newblock \enquote{Model selection in the average of inconsistent data: an
  analysis of the measured Planck-constant values.}
\newblock {\em Metrologia\/}, 49(4): 492--500.
\endbibitem

\bibitem[{Neyman and Scott(1948)}]{Neyman-Scott:1948}
Neyman, J. and Scott, E.~L. (1948).
\newblock \enquote{Consistent Estimates Based on Partially Consistent
  Observations.}
\newblock {\em Econometrica\/}, 16(1): 1--32.
\endbibitem

\bibitem[{Rao(1945)}]{Rao:1945}
Rao, C.~R. (1945).
\newblock \enquote{Information and the accuracy attainable in the estimation of
  statistical parameters.}
\newblock {\em Bull.\ Calcutta Math. Soc.\/}, 37: 81--89.
\endbibitem

\bibitem[{Samworth(2012)}]{Samworth:2012}
Samworth, R.~J. (2012).
\newblock \enquote{Stein’s Paradox.}
\newblock {\em Eureka\/}, 62: 38--41.
\endbibitem

\bibitem[{Sivia and Skilling(2006)}]{Sivia:2006}
Sivia, D. and Skilling, J. (2006).
\newblock {\em Data Analysis: A Bayesian Tutorial\/}.
\newblock Oxford: Oxford University Press.
\endbibitem

\bibitem[{Stein(1959)}]{Stein:1959}
Stein, C. (1959).
\newblock \enquote{An Example of Wide Discrepancy Between Fiducial and
  Confidence Intervals.}
\newblock {\em Ann. Math. Statist.\/}, 30(4): 877--880.
\endbibitem

\bibitem[{Toman et~al.(2012)Toman, Fischer, and Elster}]{Toman:2012}
Toman, B., Fischer, J., and Elster, C. (2012).
\newblock \enquote{Alternative analyses of measurements of the Planck
  constant.}
\newblock {\em Metrologia\/}, 49(4): 567--571.
\endbibitem

\bibitem[{von~der Linden et~al.(2014)von~der Linden, Dose, and von
  Toussaint}]{Linden:2014}
von~der Linden, W., Dose, V., and von Toussaint, U. (2014).
\newblock {\em Bayesian Probability Theory: Applications in the Physical
  Sciences\/}.
\newblock Cambridge: Cambridge University Press.
\endbibitem

\bibitem[{{Wolfram Research Inc.}(2012)}]{Mathematica:9}
{Wolfram Research Inc.} (2012).
\newblock {\em Mathematica\/}.
\newblock Version 9.0. Champaign, Illinois: Wolfram Research, Inc.
\endbibitem

\end{thebibliography}

\begin{acknowledgement}
This work was jointly funded by the European Metrology Research Programme (EMRP) participating countries within the European Association of National Metrology Institutes (EURAMET), the European Union, and the Italian ministry of education, university, and research (awarded project P6-2013, implementation of the new SI).
\end{acknowledgement}

\end{document}